\documentclass[aps,prd,eqsecnum,preprint,tightenlines,nofootinbib]{revtex4-2}
\usepackage{graphicx,latexsym}

\def\bea{\begin{eqnarray}}
\def\eea{\end{eqnarray}}
\def\be{\begin{equation}}
\def\ee{\end{equation}}
\newcommand{\ub}[1]{\underline{#1}}

\newcommand{\Pminus}{{\cal P}^-}
\newcommand{\Pplus}{{\cal P}^+}
\def\senk#1{\vec{#1}_\perp}

\begin{document}

\title{Nonperturbative light-front effective potential for static sources
in quenched scalar Yukawa theory
}
\author{Sophia S. Chabysheva}
\affiliation{Department of Physics, University of Idaho, Moscow ID 83844 USA}
\author{John R. Hiller}
\affiliation{Department of Physics, University of Idaho, Moscow ID 83844 USA}
\affiliation{Department of Physics and Astronomy,
University of Minnesota-Duluth,
Duluth, Minnesota 55812 USA}

\date{\today}

\begin{abstract}
We compute an effective potential between two fixed sources in light-front
quantization of a quenched scalar Yukawa theory that models 
the interaction of complex scalar fields through the exchange of
a neutral scalar.  Despite the breaking of explicit rotational
symmetry by the use of light-front coordinates, the effective
potential is rotationally symmetric and matches the standard
Yukawa potential for scalar exchange.  The neutral scalar field is
represented by a coherent state, which is obtained nonperturbatively
as an eigenstate of our model Hamiltonian, with the eigenenergy determining 
the effective potential.  The sources are represented by wave packets 
that are fixed with respect to ordinary time, but
move in light-front coordinates.  The theory is quenched, to
remove pair-production processes that would otherwise cause the 
spectrum to be unbounded from below.
\end{abstract}


\maketitle

\section{Introduction}   \label{sec:introduction}

A key quantity to obtain from quantum chromodynamics is the
effective potential between a quark and an 
antiquark~\cite{staticQCD}.\footnote{For recent discussion of
static sources, see, for example, \protect\cite{recentQCD}.}
This has been studied quite carefully in the lattice formulation of
QCD~\cite{lattice}.  For comparison, it would be useful to be able to do
the equivalent calculation in a nonperturbative light-front 
formulation.\footnote{For reviews of light-front
quantization and applications, 
see \protect\cite{LFreview1,LFreview2,LFreview3,LFreview4,WhitePaper,LFreview5}.} 
In order to develop a method for doing so, we study static sources in
a quenched scalar Yukawa theory, where the complications of gauge
fields and intrinsic quantum numbers can be neglected.  We focus
on formulating an eigenvalue problem that yields the energy of a
state with two static sources dressed by a cloud of scalar
particles.  While doing so, we accommodate the later necessity
of associating dynamics with intrinsic quantum numbers of the
sources.  This is to allow for sources with spin and color
charge that will change when interacting with the surrounding
cloud.  Here we restrict the model to complex source fields
with ordinary charge; however, we accommodate the possibility
of dynamical properties by placing the static sources in the quantum state,
rather than using delta-function currents in the Lagrangian.

The analysis of static sources on a light front has been considered previously.
In particular, Rozowsky and Thorn~\cite{Thorn} studied the
force between two sources on a light front by arranging
a purely transverse separation, a limitation that we avoid.
Burkardt and Klindworth~\cite{Klindworth} applied a transverse 
lattice approach~\cite{transverselattice} in (2+1)-dimensional 
QCD to the calculation of a $Q\bar{Q}$ potential which is
nearly rotationally invariant.  Blunden {\em et al}.~\cite{Blunden}
considered light-front models where a particle interacts
with a static potential.

An important aspect of our work is that we obtain the effective potential through
variation of the ordinary energy, built from light-front quantities, rather than 
the light-front energy alone.  This is the physical definition of a potential
that can be translated into a force between the sources.  The necessity of
considering the ordinary energy has been seen in other 
contexts~\cite{Vary,Elser,SDLCQ,LFCasimir}. Such a consideration is also
important because the static sources prevent momentum conservation, leaving
the ordinary energy as the only conserved part of the four-momentum.

\begin{figure}[hbt]
\vspace{0.2in}
\begin{tabular}{cc}
\includegraphics[width=7cm]{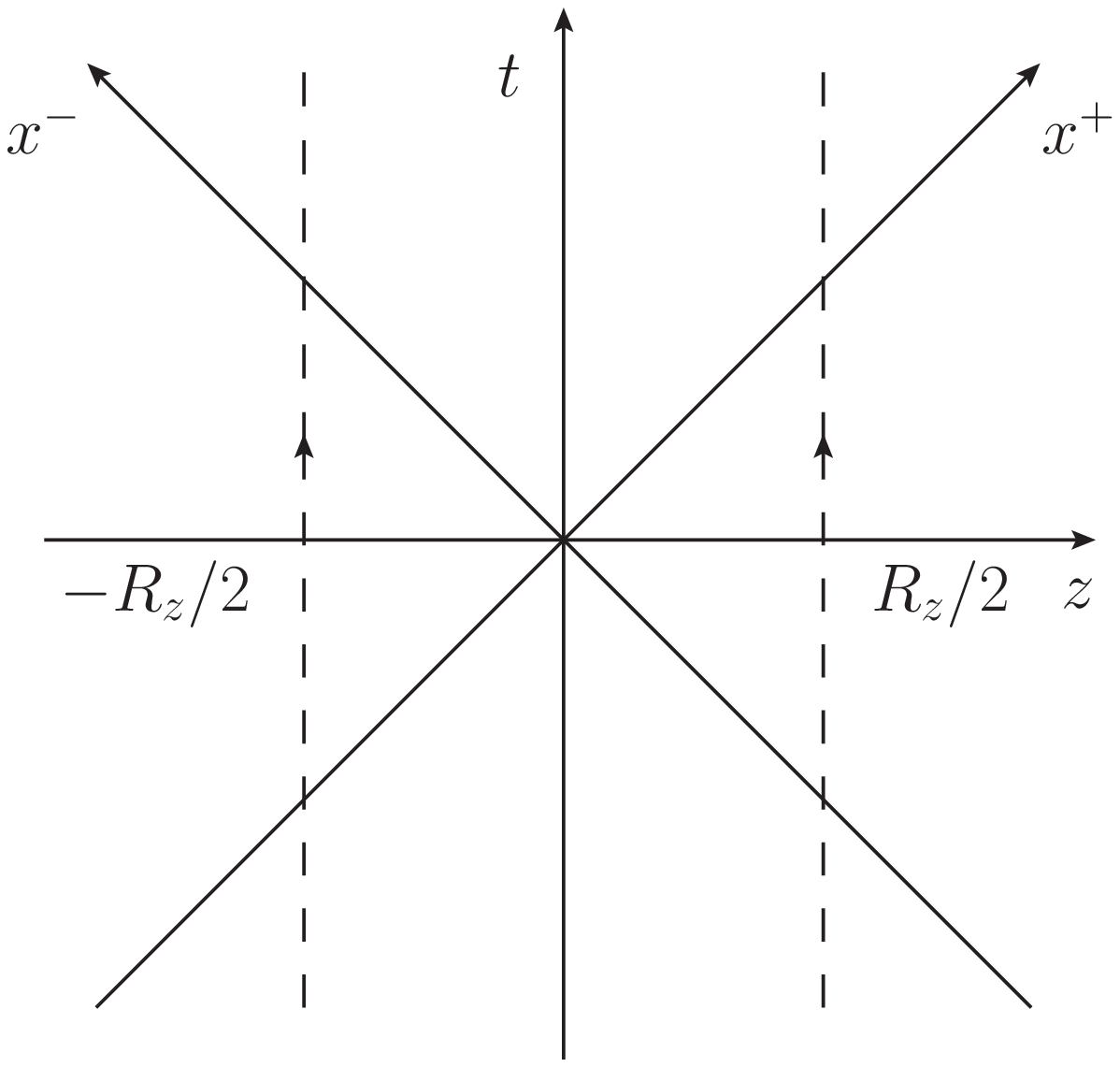} & 
\includegraphics[width=7cm]{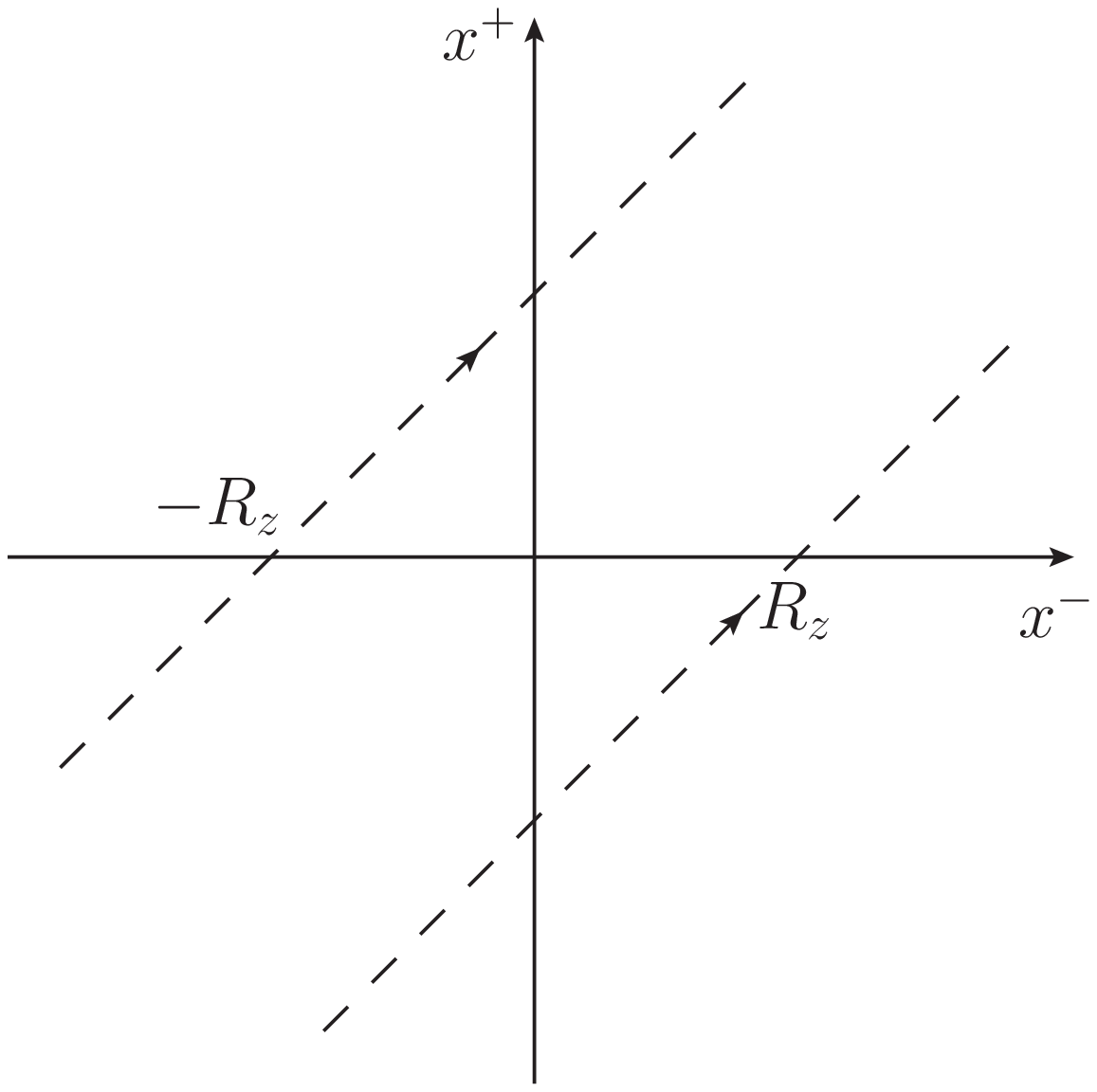} \\
(a) & (b) 
\end{tabular}
\caption{\label{fig:frames}
Static source trajectories in the $z$-$t$ plane for (a) ordinary and 
(b) light-front coordinates.
}
\end{figure}

We take the following as our definition of light-front 
coordinates~\cite{Dirac}:
\be
x^\pm\equiv (t\pm z),\;\;\senk{x}=(x,y),\;\;\ub{x}=(x^-,\senk{x}),
\ee
with $x^+$ chosen as the light-front time.
The inversion is obviously $t=\frac12(x^++x^-)$ and $z=\frac12(x^+-x^-)$.
These relationships are illustrated in Fig.~\ref{fig:frames}(a).\footnote{The figure
was drawn with JaxoDraw~\protect\cite{JaxoDraw}.}  The associated derivatives are
\be
\frac{\partial}{\partial x^\pm}=\frac12\left(\frac{\partial}{\partial t}\pm\frac{\partial}{\partial z}\right).
\ee
The light-front energy and momentum are
\be
p^-\equiv E+p_z, \;\; p^+\equiv E-p_z,\;\; \senk{p}=(p_x,p_y).
\ee
The dot product of four-vectors is $p\cdot x=\frac12 p^-x^+ +\ub{p}\cdot\ub{x}$ with
$\ub{p}\cdot\ub{x}=\frac12 p^+x^--\senk{p}\cdot\senk{x}$ as the dot product of three-vectors.
The mass-shell condition is then $p\cdot p=p^+p^--p_\perp^2=m^2$, which implies 
$p^-=(m^2+p_\perp^2)/p^+$.

In what follows, we first define quenched scalar Yukawa theory in light-front
quantization, in Sec.~\ref{sec:QSY}.  
The construction of a single static source dressed by neutral scalars
is described in Sec.~\ref{sec:single}; this includes mass
renormalization such that the eigenenergy is equal to the
physical mass of the complex scalar field.  Two static sources
are then constructed as a product of single sources placed
with fixed separation in Sec.~\ref{sec:double}; here
we obtain the key result that the energy of the two-source
eigenstate is shifted from twice the physical mass of one by
an amount equal to the Yukawa potential.  We also compute the
change in the average number of neutral scalars.  A brief summary is
given in Sec.~\ref{sec:summary}.
Details of the construction of wave packets on a light front are given in 
Appendix~\ref{sec:packets}, to properly represent a static source 
that moves in the $x^-$ direction~\cite{Blunden}.  Some aspects of the 
calculations and an alternate approach to the single-source case
are placed in additional appendices.

\section{Quenched scalar Yukawa theory} \label{sec:QSY}

Scalar Yukawa theory couples a complex scalar field $\chi$ with bare mass $m_0$ to
a real scalar field $\phi$ with mass $\mu$; the physical mass
of the complex scalar will be written as $m$. The Lagrangian is
\be
{\cal L}=\partial_\mu\chi^*\partial^\mu\chi-m_0^2|\chi|^2
  +\frac12(\partial_\mu\phi)^2-\frac12 \mu^2\phi^2-g\phi|\chi|^2.
\ee
This model is also known as the (massive) Wick--Cutkosky 
model~\cite{WickCutkosky} and has received considerable attention 
in light-front quantization~\cite{Sawicki,JiFurnstahl,WC1+1,Wivoda,Swenson,Ji:1994zx,Cooke:2000ef,%
EarlierQSY,HwangKarmanov,Karmanov,JiTokunaga} in both two and four 
dimensions.  The most recent work focuses on the 
construction of the eigenstate for a charged scalar dressed by
a cloud of neutrals~\cite{QSY,LFCCqsy}.

The quenched form of the theory excludes pair production.  Without
this restriction, the theory is ill defined, with a spectrum that
is unbounded from below, as happens in any cubic scalar theory~\cite{Baym,Gross}. 
The quenching also means that the neutral scalar does not require
mass renormalization, hence the use of the physical mass $\mu$ in
the Lagrangian.

The light-front Hamiltonian density is~\cite{LFreview5}
\be
{\cal H}=|\senk{\partial}\chi|^2+ m_0^2|\chi|^2
   +\frac12(\senk{\partial}\phi)^2+\frac12 \mu^2\phi^2+g\phi|\chi|^2.
\ee
The mode expansions for the fields are
\be
\phi(x)=\int \frac{dp^+d^2p_\perp}{\sqrt{16\pi^3 p^+}}
        \left[a(\ub{p})e^{-ip\cdot x}+a^\dagger(\ub{p})e^{ip\cdot x}\right]
\ee
and
\be
\chi(x)=\int \frac{dp^+d^2p_\perp}{\sqrt{16\pi^3 p^+}}
        \left[c_+(\ub{p})e^{-ip\cdot x}+c_-^\dagger(\ub{p})e^{ip\cdot x}\right].
\ee
The nonzero commutators are 
\be \label{eq:scalarcommutators}
{[}a(\ub{p}),a^\dagger(\ub{p'})]=\delta(\ub{p}-\ub{p'})
             \equiv\delta(p^+-p^{\prime +})\delta(\senk{p}-\senk{p'}), \;\;
{[}c_\pm(\ub{p}),c_\pm^\dagger(\ub{p'})]=\delta(\ub{p}-\ub{p'}).
\ee
The light-front Hamiltonian is $\Pminus=\Pminus_0+\Pminus_{\rm int}$,
with
\be \label{eq:Pminus0}
\Pminus_0=\int d\ub{p}\frac{m_0^2+\senk{p}^2}{p^+}
     \left[c_+^\dagger(\ub{p})c_+(\ub{p})+c_-^\dagger(\ub{p})c_-(\ub{p})\right]
     +\int d\ub{q}\frac{\mu^2+\senk{q}^2}{q^+}a^\dagger(\ub{q})a(\ub{q}),
\ee
and
\bea \label{eq:Pminusint}
\Pminus_{\rm int}&=&g\int\frac{d\ub{p} d\ub{q}}{\sqrt{16\pi^3 p^+ q^+(p^++q^+)}}
  \left[\left(c_+^\dagger(\ub{p}+\ub{q})c_+(\ub{p})
                       +c_-^\dagger(\ub{p}+\ub{q})c_-(\ub{p})\right)a(\ub{q}) \right. \\
  && \rule{2in}{0mm} \left.
     +a^\dagger(\ub{q})\left(c_+^\dagger(\ub{p})c_+(\ub{p}+\ub{q})
                            +c_-^\dagger(\ub{p})c_-(\ub{p}+\ub{q})\right)\right] 
        \nonumber \\
 && +g\int \frac{d\ub{p}_1 d\ub{p}_2}{\sqrt{16\pi^3 p_1^+ p_2^+ (p_1^++p_2^+)}}
\left[c_+^\dagger(\ub{p}_1)c_-^\dagger(\ub{p}_2)a(\ub{p}_1+\ub{p}_2)
      +a^\dagger(\ub{p}_1+\ub{p}_2)c_+(\ub{p}_1)c_-(\ub{p}_2)\right]. \nonumber
\eea
For the quenched theory, the second term in $\Pminus_{\rm int}$ is dropped.
The light-front momentum operator is
\be \label{eq:Pplus}
\Pplus=\int d\ub{q}\, q^+ a^\dagger(\ub{q})a(\ub{q})
  +\int d\ub{p} \,p^+ [c_+^\dagger(\ub{p})c_+(\ub{p})+c_-^\dagger(\ub{p})c_-(\ub{p})].
\ee
We then define an ordinary energy operator as
\be
{\cal E}=\frac12(\Pminus+\Pplus).
\ee
This will play a key role, because momentum is not conserved when static sources
are present.

\section{Single source}  \label{sec:single}

\subsection{Wave packet for the source}

A static source is not at rest in light-front coordinates.  It
moves steadily in the positive $x^-$ direction, as indicated in 
Fig.~\ref{fig:frames}(b). For this we consider light-front
wave packets for the sources, which are discussed in detail 
in Appendix~\ref{sec:packets}.
We place a single source at $\pm\vec{R}/2$ with a state given by
\be
|F^\pm\rangle=\int d\ub{p}  \sqrt{p^+}F^\pm(\ub{p})c_\pm^\dagger(\ub{p})|0\rangle,
\ee
where $F^\pm(\ub{p})$ is a momentum-space envelope function peaked at a light-front
momentum for an object of mass $m$ at rest, $\ub{p}=(m,\vec{0}_\perp)$. 
The explicit factor of $\sqrt{p^+}$ is included to facilitate
calculation of expectation values for the light-front energy and 
the probability density. 

The model is intended to produce a spatial probability density that
is sharply peaked at the source locations $\pm\vec{R}/2$.  To
impose this, we compute the expectation value of $|\chi|^2$
for the state $|F^\pm\rangle$
\bea
\langle F^\pm|\!:\!|\chi^2|\!:\!|F^\pm\rangle
&=&\langle F^\pm|\int\frac{d\ub{p}'}{\sqrt{16\pi^3p^{\prime +}}}\frac{d\ub{p}}{\sqrt{16\pi^3p^+}}
\left[c_+^\dagger(\ub{p}')c_+(\ub{p})e^{i(p'-p)\cdot x} \right.  \\
&& \rule{1in}{0mm} \left. +c_-^\dagger(\ub{p}')c_-(\ub{p})e^{i(p'-p)\cdot x}+\cdots\right]|F^\pm\rangle.
\nonumber
\eea
Here the extra dots indicate terms that do not contribute.  Contractions of the various
operators yield delta functions that resolve all but two sets of integrals.  We 
evaluate at $x^+=0$:
\be
\langle F^\pm|\!:\!|\chi^2|\!:\!|F^\pm\rangle|_{x^+=0}
=\int\frac{d\ub{p}'}{\sqrt{16\pi^3}}F^{\pm*}(\ub{p}')e^{i\ub{p}'\cdot\ub{x}}
  \int\frac{d\ub{p}}{\sqrt{16\pi^3}}F^\pm(\ub{p})e^{-i\ub{p}\cdot\ub{x}}.
\ee
With the definition
\be
\psi^\pm(\ub{x})=\int\frac{d\ub{p}}{\sqrt{16\pi^3}}F^\pm(\ub{p})e^{-i\ub{p}\cdot\ub{x}},
\ee
the expectation value reduces to 
\be  \label{eq:chi2}
\langle F^\pm|\!:\!|\chi^2|\!:\!|F^\pm\rangle|_{x^+=0}=|\psi^\pm(\ub{x})|^2.
\ee

To represent a static source, we require that this become a delta function
when the spatial packet becomes infinitesimally narrow\footnote{The constraint 
for $x^-$ is obtained by requiring $z=\frac12(x^+-x^-)=\pm R_z/2$ to hold at
light-front time $x^+=0$.}
\be \label{eq:spatialdelta}
\langle F^\pm|\!:\!|\chi^2|\!:\!|F^\pm\rangle|_{x^+=0}
=N^2\delta(x^-\pm R_z)\delta(\senk{x}\mp\senk{R}/2),
\ee
with $N^2$ a normalization factor to be determined. Therefore, we identify
\be  \label{eq:psidelta}
|\psi^\pm(\ub{x})|^2\rightarrow N^2 \delta(x^-\pm R_z)\delta(\senk{x}\mp\senk{R}/2).
\ee
The momentum-space wave-packet envelope is the Fourier transform
\be
F^\pm(\ub{p})=\int \frac{d\ub{x}}{\sqrt{16\pi^3}}e^{i\ub{p}\cdot\ub{x}}\psi^\pm(\ub{x}).
\ee
As transforms, $\psi^\pm$ and $F^\pm$ share the common normalization
\be
N^2=\int d\ub{x} |\psi^\pm(\ub{x})|^2=\int d\ub{p} |F^\pm(\ub{p})|^2,
\ee
where the integral over $p^+$ can be extended to $-\infty$ because the envelope 
$F^\pm$ is sharply peaked about $p^+=m>0$.  

The value of the normalization 
for $F^\pm$ is separately determined by the normalization of $|F^\pm\rangle$ as
\be
1=\langle F^\pm|F^\pm\rangle=\int d\ub{p}\, p^+|F^\pm(\ub{p})|^2=m\int d\ub{p} |F^\pm(\ub{p})|^2=mN^2,
\ee
where $p^+$ is replaced by the peak value of $m$ and the $p^+$ integration
is again extended to $-\infty$.  Thus, we have $N=1/\sqrt{m}$, which is
dimensionally consistent with the Hamiltonian term $m^2|\chi|^2$ being an
energy density.

The expectation value for the free part of the light-front 
energy for the complex field is given by
\be \label{eq:<P-chi>}
\langle \Pminus_{0\chi}\rangle=\int d\ub{x} \langle F^\pm|\!:\!|\senk{\partial}\chi|^2+m_0^2|\chi|^2:\!|F^\pm\rangle.
\ee
The first term can be evaluated from
\be
\senk{\partial}\chi=\int\frac{d\ub{p}}{\sqrt{16\pi^3 p^+}}
     i\senk{p}[c_+(\ub{p})e^{-ip\cdot x}-c_-^\dagger(\ub{p})e^{ip\cdot x}].
\ee
The presence of the $\senk{p}$ factor means that the first term in (\ref{eq:<P-chi>})
is zero, because the wave packet is symmetrically peaked at $\senk{p}=0$.
The second term is readily obtained from (\ref{eq:spatialdelta}),
with the normalization $N^2=1/m$, as
\be
\int d\ub{x} \langle F^\pm|\!:\!|m_0^2|\chi|^2:\!|F^\pm\rangle
=\frac{m_0^2}{m}\int d\ub{x} \delta(x^-\pm R_z)\delta(\senk{x}\mp\senk{R}/2)
=\frac{m_0^2}{m}.
\ee
The expectation value for the 
light-front longitudinal momentum is $m$, because the wave packets
are sharply peaked at $p^+=m$.  Thus for ${\cal E}=(\Pminus+\Pplus)/2$ we have
\be
 \langle F^\pm|{\cal E}|F^\pm\rangle=m_0^2/2m+m/2,
\ee
which reduces to $m$ when the coupling $g$ is zero and the mass is not renormalized.
The presence of interactions with the neutral scalar will renormalize
the mass, as we show below. 

\subsection{Coherent state for the neutrals}

On this source state $|F^\pm\rangle$, we build a coherent state of neutrals 
as an {\em ansatz} for the solution
\be
|G_1^\pm F^\pm\rangle=\sqrt{Z_1^\pm}e^{\int d\ub{q}G_1^\pm(\ub{q})a^\dagger(\ub{q})}|F^\pm\rangle,
\ee
with $\sqrt{Z_1^\pm}$ a normalization factor given by
\be
Z_1^\pm=e^{-\int d\ub{q}|G_1^\pm(\ub{p})|^2}.
\ee
It is an eigenstate of the annihilation operator
\be
a(\ub{q})|G_1^\pm F^\pm\rangle=G_1^\pm(\ub{q})|G_1^\pm F^\pm\rangle,
\ee
and we require it also to be an eigenstate of the energy 
operator ${\cal E}$
\be
{\cal E}|G_1^\pm F^\pm\rangle=E^\pm|G_1^\pm F^\pm\rangle.
\ee
This eigenvalue condition, projected onto $\langle F^\pm|$, reduces to
\bea  \label{eq:EG1}
\left[\frac{m_0^2}{2m}\right.&+&\left.\frac12 m\right]|G_1^\pm\rangle
+\int d\ub{q} \frac12\left[\frac{q_\perp^2+\mu^2}{q^+}+q^+\right] a^\dagger(\ub{q})G_1^\pm(\ub{q})|G_1^\pm\rangle \\
    &&    +\frac{g}{2m}\int \frac{d\ub{q}}{\sqrt{16\pi^3 q^+}}
        \left\{ e^{\pm iq^+R_z/2\pm i\senk{q}\cdot\senk{R}/2} G_1^\pm(\ub{q})
           +e^{\mp iq^+R_z/2\mp i\senk{q}\cdot\senk{R}/2}a^\dagger(\ub{q})\right\}|G_1^\pm\rangle \nonumber \\
    &&  =E^\pm|G_1^\pm\rangle,  \nonumber
\eea
where we have used (\ref{eq:spatialdelta}) to replace $\langle F^\pm||\chi|^2|F^\pm\rangle$ and used the
resulting delta functions to do the $\ub{x}$ integrals in $\Pminus_{\rm int}=\int d\ub{x}\, g\phi|\chi|^2$.

For this eigenvalue equation to be satisfied, the terms that contain $a^\dagger$ must cancel,
which means that
\be \label{eq:G1eqn}
\frac12\left[ \frac{q_\perp^2+\mu^2}{q^+} + q^+\right] G_1^\pm(\ub{q})
    +\frac{g}{2m}\frac{1}{\sqrt{16\pi^3 q^+}}
         e^{\mp iq^+R_z/2\mp i\senk{q}\cdot\senk{R}/2}=0.
\ee
The function $G_1^\pm$ must then be
\be  \label{eq:G1}
G_1^\pm(\ub{q})=-\frac{g}{m}\sqrt{\frac{q^+}{16\pi^3}}
      \frac{e^{\mp iq^+R_z/2\mp i\senk{q}\cdot\senk{R}/2}}
         {(q^+)^2+q_\perp^2+\mu^2}.
\ee

The eigenenergy is then computed from the remaining terms as
\bea
E^\pm&=&\frac{m_0^2}{2m}+\frac12 m +\frac{g}{2m}\int \frac{d\ub{q}}{\sqrt{16\pi^3 q^+}}
       e^{\pm iq^+R_z/2\pm i\senk{q}\cdot\senk{R}/2} G_1^\pm(\ub{q}) \nonumber\\
      &=&\frac{m_0^2}{2m}+\frac12 m -\frac12\left(\frac{g}{m}\right)^2
      \int \frac{d\ub{q}}{16\pi^3}  \frac{1}{(q^+)^2+q_\perp^2+\mu^2}. 
\eea
The integral in the last term is divergent.  We introduce a cutoff $\Lambda$
and define
\be \label{eq:ILambda}
I(\Lambda)= \int \frac{d\ub{q}}{16\pi^3\mu} \frac{\theta(\Lambda^2-(q^+)^2-q_\perp^2)}{(q^+)^2+q_\perp^2+\mu^2},
\ee
and the eigenenergy becomes
\be
E^\pm=\frac{m_0^2}{2m}+\frac12 m-\frac12\left(\frac{g}{m}\right)^2 \mu I(\Lambda).
\ee
We can arrange $E^\pm=m$, the physical mass, by choosing
the bare mass such that
\be  \label{eq:baremass}
m_0^2=m^2+g^2\frac{\mu}{m}I(\Lambda).
\ee
The cutoff dependence is then removed.

We now have an exact solution for the single-source state that includes the
source and a Fock-state expansion in the number of neutral scalars.  The form is independent of
the source location, except for a phase in the individual wave functions
$G_1^\pm$. We also see that, in both cases, the normalization of
the state is determined by 
\be
Z_1^\pm=e^{-\left(\frac{g}{m}\right)^2\int \frac{d\ub{q}}{16\pi^3}\frac{q^+}{((q^+)^2+q_\perp^2+\mu^2)^2}}.
\ee
The integral is divergent and requires a cutoff, to give meaning to
the norm of the state.

Assuming that such a cutoff is in place, the single-source problem can
be also formulated as a variational problem
with $|G_1^\pm F^\pm\rangle$ as the trial state. The expectation
value of the energy is
\bea
\langle G_1^\pm F^\pm|\!:\!{\cal E}\!:\!|G_1^\pm F^\pm\rangle
&=&\frac{m_0^2}{2m}+\frac12 m
+\int d\ub{q} \frac12\left[\frac{q_\perp^2+\mu^2}{q^+}+q^+\right] G_1^{\pm *}(\ub{q})G_1^\pm(\ub{q}) \\
&&+\frac{g}{m}\int\frac{d\ub{q}}{\sqrt{16\pi^3 q^+}}
  \left[e^{\pm iq^+R_z/2\pm i\senk{q}\cdot\senk{R}/2} G_1^\pm(\ub{q})
           +e^{\mp iq^+R_z/2\mp i\senk{q}\cdot\senk{R}/2}G_1^{\pm *}(\ub{q})\right]. \nonumber
\eea
Variation with respect to $G_1^{\pm *}$ yields (\ref{eq:G1eqn}) and the
same results follow, including the value $m$ of $\langle G_1^\pm F^\pm|\!:\!{\cal E}\!:\!|G_1^\pm F^\pm\rangle$
at this minimum, once the mass renormalization is taken into account.

For comparison with the double-source state, we also consider the 
average number $\langle n\rangle_\pm$ of neutral scalars in this single-source state.  This
is computed from the coherent state as
\be
\langle n\rangle_\pm\equiv\int d\ub{q}\langle G_1^\pm F^\pm|a^\dagger(\ub{q})a(\ub{q})|G_1^\pm F^\pm\rangle
     =\int d\ub{q}|G_1^\pm(\ub{q})|^2.
\ee
However, on substitution of the form for $G_1^\pm$, the expectation value reduces to
\be
\langle n\rangle_\pm=\left(\frac{g}{m}\right)^2\int \frac{d\ub{q}}{16\pi^3} \frac{(q^+)^2}{[(q^+)^2+q_\perp^2+\mu^2]^2},
\ee
which contains the same divergent integral that defines the normalization.  Thus, unless one chooses to
fix the coupling $g$ as a bare coupling by fixing the value of $\langle n\rangle_\pm$, the
number of neutrals that dress a single source is effectively infinite.  Instead of invoking
a restriction on the coupling, we will accept this infinite value as part
of the nature of the single-source state and investigate the change in 
the number of neutrals induced by the presence of a second source.

\section{Double source} \label{sec:double}

The case of two static sources can also be solved with a coherent state.  We take 
the variational approach, with a trial state
built from a product of single-source states of opposite charge
\be
|G_2G_1^+G_1^-F^+F^-\rangle=\sqrt{\frac{Z}{Z_1^+Z_1^-}}
     e^{\int d\ub{q}G_2(\ub{q})a^\dagger(\ub{q})}|G_1^+F^+\rangle|G_1^-F^-\rangle,
\ee
where $G_2(\ub{q})$ is a function to be determined and
$Z$ fixes the overall normalization as
\be
Z=e^{-\int d\ub{q}\,|G_2(\ub{q})+G_1^+(\ub{q})+G_1^-(\ub{q})|^2}.
\ee
It will turn out that $G_2(\ub{q})$ is actually zero.

We choose the two sources to be of opposite charge, to have a simpler
calculation without the cross terms that would arise for two identical
sources with the same charge.  However, this is not a serious restriction because one
can argue that the overlap between spatial packets is effectively zero.
This would allow the cross terms to be ignored.

When there are two sources, the expectation value of $:\!|\chi|^2:$ becomes
\bea
\lefteqn{\langle F^+F^-|\!:\!|\chi^2|\!:\!|F^+F^-\rangle|_{x^+=0}}  &&  \\
&& =\int\frac{d\ub{p}'}{\sqrt{16\pi^3p^{\prime +}}}F^{+*}(\ub{p}')e^{i\ub{p}'\cdot\ub{x}}
  \int\frac{d\ub{p}}{\sqrt{16\pi^3p^+}}F^+(\ub{p})e^{-i\ub{p}\cdot\ub{x}}
    \int d\ub{p}_2 p_2^+|F^-(\ub{p}_2)|^2 \nonumber \\
&&  \rule{1in}{0mm} +\int\frac{d\ub{p}'}{\sqrt{16\pi^3p^{\prime +}}}F^{-*}(\ub{p}')e^{i\ub{p}'\cdot\ub{x}}
    \int\frac{d\ub{p}}{\sqrt{16\pi^3p^+}}F^-(\ub{p})e^{i\ub{p}\cdot\ub{x}}
  \int d\ub{p}_1 p_1^+|F^+(\ub{p}_1)|^2. \nonumber
\eea
With the normalization requirement $\int d\ub{p}\, p^+|F^\pm(\ub{p})|^2=1$
and the limit (\ref{eq:psidelta}), this reduces to
\be \label{eq:deltafnlimit}
\langle F^+F^-|\!:\!|\chi^2|\!:\!|F^+F^-\rangle|_{x^+=0}
=\frac{1}{m}[\delta(x^-+ R_z)\delta(\senk{x}-\senk{R}/2)+\delta(x^-- R_z)\delta(\senk{x}+\senk{R}/2)],
\ee
which places the sources appropriately at $\vec{R}/2$ and $-\vec{R}/2$.
In terms of light-front coordinates, 
the peaks are at $x^-=\mp R_z$ and $\senk{x}=\pm\senk{R}/2$ when $x^+=0$.

The expectation value for ${\cal E}$ in the state is
\bea  \label{eq:initial<E>}
\langle :\!{\cal E}\!:\rangle&=&\frac12\left[2\frac{m_0^2}{m}+2m\right]
+\int d\ub{q}\frac12\left[\frac{q_\perp^2+\mu^2}{q^+}+q^+\right]\left|G_2(\ub{q})+G_1^+(\ub{q})+G_1^-(\ub{q})\right|^2 \\
&&+\frac{g}{2m}\int\frac{d\ub{q}}{\sqrt{16\pi^3 q^+}}
    \left\{e^{iq^+R_z/2+i\senk{q}\cdot\senk{R}/2}+e^{-iq^+R_z/2-i\senk{q}\cdot\senk{R}/2}\right\} \nonumber \\
 && \rule{1in}{0mm} \times\left[G_2(\ub{q})+G_1^+(\ub{q})+G_1^-(\ub{q})+G_2^*(\ub{q})+G_1^{+*}(\ub{q})+G_1^{-*}(\ub{q})\right].
    \nonumber
\eea
Variation with respect to $G_2^*$ results in 
\be
\frac12\left[\frac{q_\perp^2+\mu^2}{q^+}+q^+\right]\left[G_2(\ub{q})+G_1^+(\ub{q})+G_1^-(\ub{q})\right]
  +\frac{g}{2m}\frac{1}{\sqrt{16\pi^3 q^+}}
    \left\{e^{iq^+R_z/2+i\senk{q}\cdot\senk{R}/2}+{\rm c.c.}\right\}=0.
\ee
Substitution of the known expressions for $G_1^\pm(\ub{q})$ leaves $G_2(\ub{q})=0$.  In
other words, the effect of combining two single sources is included in
the phase associated with each source location and the interference terms
between $G_1^+$ and $G_1^-$.

On use of $G_2=0$ and the expressions for $G_1^\pm$, the expectation
value $\langle:\!{\cal E}\!:\rangle$ becomes, as discussed in Appendix~\ref{sec:<E>},
\be
\langle :\!{\cal E}\!:\rangle=\frac{m_0^2}{m}+m -\left(\frac{g}{m}\right)^2\mu I(\Lambda)
   -\left(\frac{g}{2m}\right)^2\frac{e^{-\mu R}}{4\pi R},
\ee
The first three terms of $\langle :\!{\cal E}\!:\rangle$ reduce to $2m$, with use of 
the constraint (\ref{eq:baremass}) for the bare mass. This brings us to our key result
\be
\langle :\!{\cal E}\!:\rangle=2m-\left(\frac{g}{2m}\right)^2\frac{e^{-\mu R}}{4\pi R},
\ee
which shows that the eigenenergy of two static sources is their total mass
plus a rotationally symmetric, attractive Yukawa potential in the
standard form for scalar exchange between scalars.\footnote{This differs slightly
from the Yukawa potential between fermions, because in that case
the interaction term in the Lagrangian is $g\phi\bar{\psi}\psi$
and $g$ is dimensionless; here $g$ has units of mass.}

Perhaps the most remarkable aspect is the rotational symmetry, despite the
explicit breaking of rotational symmetry by light-front coordinates.  This
is achieved not by fine tuning but by staying close to the physics of the 
configuration, in that the effective potential between the sources is 
contained within the ordinary energy not the light-front energy and
the sources are static with respect to ordinary time not light-front time.

The change in the number of neutral scalars, induced by the proximity
of the two sources, is given by
\bea
\langle\delta n\rangle&\equiv& \int d\ub{q} \langle a^\dagger(\ub{q})a(\ub{q})\rangle
                                     -\langle n\rangle_+-\langle n\rangle_- \\
  &=&\int d\ub{q} |G_1^+(\ub{q})+G_1^-(\ub{q})|^2-\int d\ub{q} |G_1^+(\ub{q})|^2-\int d\ub{q} |G_1^-(\ub{q})|^2.
\eea
Again, only the interference terms contribute
\be
\langle\delta n\rangle=\int d\ub{q} \left[G_1^{+*}(\ub{q})G_1^-(\ub{q})+G_1^{-*}(\ub{q})G_1^+(\ub{q})\right].
\ee
Substitution of the form for $G_1^\pm$ and some additional calculus, shown
in Appendix~\ref{sec:deltan}, reduces this to
\be
\langle\delta n\rangle=-\frac{1}{16\pi^2}\left(\frac{g}{m}\right)^2
\left[e^{\mu R}{\rm Ei}(-\mu R)+e^{-\mu R}{\rm Ei}(\mu R)\right],
\ee
where ${\rm Ei}$ is the exponential integral function~\cite{DLMF}.
For large separations $R$, this simplifies to
\be
\langle\delta n\rangle=-\frac{1}{8\pi^2}\left(\frac{g}{m}\right)^2\frac{1}{(\mu R)^2}+{\cal O}(\frac{1}{(\mu R)^3}),
\ee
which correctly goes to zero as the separation becomes infinite.

\section{Summary} \label{sec:summary}

We have shown that, by considering the ordinary energy of static
sources fixed with respect to ordinary time, a 
light-front calculation yields the correct Yukawa potential.
Rotational symmetry is maintained despite the explicit breaking
by the light-front coordinates themselves and without fine tuning
of parameters.  The effective potential arises from
the overlap between the clouds of neutral scalars that dress
the sources.  It is essentially an interference term in the
expectation value of the energy.

The success of the calculation is due to two factors.
One is that we consider the ordinary energy $E$, not
the light-front energy $P^-$.  The other is that the sources
are fixed with respect to ordinary time, not light-front time $x^+$.
This is analogous to our work on the Casimir effect~\cite{LFCasimir}.
The primary observation is that changing coordinate systems
does not and should not change the physics.

The calculation is nonperturbative, even though the resulting
Yukawa potential is of order $g^2$.  The eigensolution is 
obtained to all orders in $g$ as a coherent state of 
neutral scalars.  Such a solution is possible because
the static sources remove the constraint of momentum
conservation.

The approach can be extended to more complicated theories,
although the solution of the eigenvalue problem will
typically require numerical techniques.  An obvious next
application is to standard Yukawa theory with two 
fermions as sources static in position but
dynamic with respect to spin.  This can be done first
as quenched but then also with fermion-pair
contributions.  Static-source potentials in QED and QCD
are also clearly of interest; we suggest that the present
work provides a starting point.


\appendix

\section{Light-front wave packets} \label{sec:packets}

To be able to incorporate the steady movement in $x^-$ of a static
source, we consider the quantum mechanics of wave packets
on a light front.  We use such wave packets for the sources
dressed by neutral scalars in a Fock-state expansion 
or a coherent state for the neutral scalars.

The ordinary time evolution of a particle with wave function
$\Psi$ is determined by the usual Schr\"odinger equation 
$i\frac{\partial\Psi}{\partial t}={\cal P}^0\Psi$.  The
action of the momentum operator ${\cal P}^z$ is represented
by $-i\frac{\partial}{\partial z}$.  Therefore, the light-front
time evolution is determined by
\be
i\frac{\partial\Psi}{\partial x^+}=\frac{i}{2}\left(\frac{\partial}{\partial t}+\frac{\partial}{\partial z}\right)\Psi
   =\frac12({\cal P}^0-{\cal P}^z)\Psi=\frac12\Pminus\Psi.
\ee

Separation of variables is then applied, with $\Psi(\ub{x},x^+)=\tau(x^+)\psi(\ub{x})$,
to find
\be
\frac{2i}{\tau}\frac{d\tau}{dx^+}=\frac{1}{\psi}\Pminus\psi\equiv \frac{m^2+p_\perp^2}{p^+},
\ee
where the separation constant is written in the form of the on-shell light-front energy $p^-$
for a particle of mass $m$ with light-front momentum $\ub{p}=(p^+,\vec{p}_\perp)$. The light-front time evolution
is then $\tau=\exp\left(-i\frac{m^2+p_\perp^2}{2p^+}x^+\right)$.

In momentum space, $\Pminus$ for a free particle of momentum $\ub{p}$
is just the multiplicative operator $\frac{m^2+p_\perp^2}{p^+}$, and
$\phi_{\ub{p}}(\ub{q})=N\delta(q^+-p^+)\delta(\senk{q}-\senk{p})$ is the
eigenfunction.  A
Fourier transform yields $\psi_{\ub{p}}(\ub{x})=\tilde{N}e^{i\ub{p}\cdot\ub{x}}$.  A wave packet,
with momentum envelope $\phi(\ub{p})$, is then given by
\be
\Psi(\ub{x},x^+)=\int\frac{d\ub{p}}{\sqrt{16\pi^3}}\phi(\ub{p})
       \exp\left[i\left(\ub{p}\cdot\ub{x}-\frac{m^2+p_\perp^2}{2p^+}x^+\right)\right].
\ee
The normalization factor contains 16 rather than 8 because the
dot product contains a factor of 1/2 for the $p^+x^-$ term.

For a static source, we have $\senk{p}=0$; however, $p^+$ is not zero.  For a source
at $z=\pm R_z/2$, we must have $\frac12(x^+-x^-)=\pm R_z/2$ or $x^-=x^+\mp R_z$.
Thus, $x^-$ increases with light-front time~\cite{Blunden}.
For the wave packet, this corresponds to the factor 
$\exp\left[i\left(\frac12 p^+ x^--\frac{m^2}{2p^+}x^+\right)\right]$, which, to
be consistent with $x^- - x^+$ being constant for the trajectory, must have $p^+=m$.
This is, of course, the usual value for $p^+$ when the particle is at rest,
but this simple analysis shows the connection with the associated wave packet
and establishes that the envelope $\phi$ must be peaked at $\ub{p}=(m,\senk{0})$.

As an example of these envelope functions, we consider a Gaussian form,
parameterized by a width $\epsilon$, that becomes a delta function in the
appropriate limit.\footnote{The analysis presented in the main sections is independent of
the specific form chosen.}  The peak momentum value of $p^+=m$ is achieved by
including a phase factor $e^{-imx^-/2}$
\be
\psi^\pm(\ub{x})=\frac{1/\sqrt{m}}{(\epsilon\sqrt{\pi})^{3/2}}e^{-imx^-/2}
                 e^{-(x^-\pm R_z)^2/2\epsilon^2}e^{-(\senk{x}\mp\senk{R}/2)^2/2\epsilon^2}.
\ee
The Fourier transform is
\be
\phi^\pm(\ub{p})=\frac{1}{\sqrt{2m}}\left(\frac{\epsilon}{\sqrt{\pi}}\right)^{3/2}
           e^{\mp i(p^+-m)R_z/2}e^{\pm i\senk{p}\cdot\senk{R}/2}
                 e^{-\epsilon^2(p^+-m)^2/8}e^{-\epsilon^2p_\perp^2/2}.
\ee
For a static source, the momentum and position distributions are both sharply 
defined.  In our units, where $\hbar=1$, this is not mathematically obvious;
the correct relationship is recovered in an $\hbar\rightarrow0$ limit.

\section{Fock-space expansion method for the single-source problem} \label{sec:alternative}


We construct an eigenstate of ${\cal E}$ as a Fock-state expansion for the
neutrals, built on a state for a single static source at $\pm\vec{R}/2$
\be
|\psi F^\pm\rangle=\sum_n\left(\prod_i^n\int d\ub{p}_i\right)\psi_n(\ub{p}_1,\ldots,\ub{p}_n)
   \frac{1}{\sqrt{n!}}\prod_i^n a^\dagger(\ub{p}_i)|F^\pm\rangle.
\ee
Because of the static source, (light-front) momentum is not conserved, and
we do not seek simultaneous eigenstates for $\Pminus$ and $\Pplus$.  Thus
the Fock-state expansion does not contain a momentum conserving delta
function to restrict the integrals over individual momenta.  We simply
require ${\cal E}|\psi F^\pm\rangle=E^\pm|\psi F^\pm\rangle$.

The action of individual parts of ${\cal E}$ 
(See Eqs.~(\ref{eq:Pminus0})-(\ref{eq:Pplus}).) yield the following
\be
\Pminus_0|\psi F^\pm\rangle=\frac{m_0^2}{m}|\psi F^\pm\rangle
  +\sum_n\left(\prod_i^n\int d\ub{p}_i\right)\left(\sum_i^n\frac{p_{\perp i}^2+\mu^2}{p_i^+}\right)
     \psi_n(\ub{p}_1,\ldots,\ub{p}_n)
   \frac{1}{\sqrt{n!}}\prod_i^n a^\dagger(\ub{p}_i)| F^\pm\rangle,
\ee
\be
\Pplus|\psi F^\pm\rangle=m|\psi F^\pm\rangle
  +\sum_n\left(\prod_i^n\int d\ub{p}_i\right)\left(\sum_i^n\,p_i^+\right)
     \psi_n(\ub{p}_1,\ldots,\ub{p}_n)
   \frac{1}{\sqrt{n!}}\prod_i^n a^\dagger(\ub{p}_i)| F^\pm\rangle,
\ee
and
\bea
\Pminus_{\rm int}|\psi F^\pm\rangle&=&g\sum_n\left(\prod_i^n\int d\ub{p}_i\right)\psi_n(\ub{p}_1,\ldots,\ub{p}_n)
\int d\ub{x}:\!|\chi|^2:  \\
&& \times\int\frac{d\ub{q}}{\sqrt{16\pi^3 q^+}}
     \left[a(\ub{q})e^{-iq^+x^-/2+i\senk{q}\cdot\senk{x}}
           +a^\dagger(\ub{q})e^{iq^+x^-/2-i\senk{q}\cdot\senk{x}}\right]
   \frac{1}{\sqrt{n!}}\prod_i^n a^\dagger(\ub{p}_i)| F^\pm\rangle. \nonumber
\eea
Projection of ${\cal E}|\psi F^\pm\rangle=E^\pm|\psi F^\pm\rangle$ onto
$\langle F^\pm|\frac{1}{\sqrt{n'!}}\prod_i^{n'} a(\ub{q}_i)$ yields
\bea
\lefteqn{\left[\frac{m_0^2}{2m}+\frac12 m
          +\sum_i^{n'}\frac12\left(\frac{q_{\perp i}^2+\mu^2}{q_i^+}+q_i^+\right)\right]
                                \psi_{n'}(\ub{q}_1,\ldots,\ub{q}_{n'})} && \\
&& +\frac{g}{2m}\sum_j^{n'}\frac{1}{\sqrt{16\pi^3 q^+}}
   e^{\mp iq_j^+R_z/2\mp i\vec{q}_{\perp j}\cdot\senk{R}/2}
          \frac{1}{\sqrt{n'}}\psi_{n'-1}(\ub{q}_1,\ldots,\ub{q}_{j-1},\ub{q}_{j+1},\ldots,\ub{q}_{n'}) \nonumber \\
&& +\frac{g}{2m}\sum_j^{n'+1}\int \frac{d\ub{q}}{\sqrt{16\pi^3 q^+}} 
   e^{\pm iq^+R_z/2\pm i\senk{q}\cdot\senk{R}/2}
\frac{1}{\sqrt{n'+1}}\psi_{n'+1}(\ub{q}_1,\ldots,\ub{q}_{j-1},\ub{q},\ub{q}_j,\ldots,\ub{q}_{n'}) \nonumber \\
&& = E^\pm\psi_{n'}(\ub{q}_1,\ldots,\ub{q}_{n'}).  \nonumber
\eea
An analytic solution is obtained by writing $\psi_n$ as a product
of single-particle wave functions $G_1^\pm(\ub{q}_i)$
\be \label{eq:factoredpsi}
\psi_n(\ub{q}_1,\ldots,\ub{q}_n)= \frac{1}{\sqrt{n!}}\prod_i^n G_1^\pm(\ub{q}_i).
\ee
Substitution of this product, and division by $\psi_{n'}$, leaves
\bea \label{eq:projected-divided}
\sum_i^{n'}\left[\frac12\left(\frac{q_{\perp i}^2+\mu^2}{q_i^+}+q_i^+\right)\right.
&+&\left.\frac{g}{2m}\frac{1}{\sqrt{16\pi^3 q^+}}
   \frac{e^{\mp iq_i^+R_z/2\mp i\vec{q}_{\perp i}\cdot\senk{R}/2}}{G_1^\pm(\ub{q}_i)}\right] \\
&& +\frac{m_0^2}{2m}+\frac12 m +\frac{g}{2m} \frac{1}{n'+1}\sum_j^{n'+1} f^\pm(\vec{R})= E^\pm, \nonumber
\eea
where
\be
f^\pm(\vec{R})\equiv \int \frac{d\ub{q}}{\sqrt{16\pi^3 q^+}} G_1^\pm(\ub{q})
   e^{\pm iq^+R_z/2\pm i\senk{q}\cdot\senk{R}/2}.
\ee
The equation is solved provided that the content of the square brackets is
zero and that $E^\pm$ is given by 
\be
E^\pm=\frac{m_0^2}{2m}+\frac12 m+\frac{g}{2m}f^\pm(\vec{R}).
\ee
We will find that for this single-source case
any $\vec{R}$ dependence is actually absent;
the function $f^\pm$ is simply constant.

For the square bracket in (\ref{eq:projected-divided}) to be zero,
we need
\be
G_1^\pm(\ub{q})=-\frac{g}{m}\sqrt{\frac{q^+}{16\pi^3}}
    \frac{e^{\mp iq_i^+R_z/2\mp i\vec{q}_{\perp i}\cdot\senk{R}/2}}{(q^+)^2+q_\perp^2+\mu^2}.
\ee
Substitution into the expression for $f^\pm$ and then into $E^\pm$ yields
\be
E^\pm=\frac{m_0^2}{2m}+\frac{m}{2}-\frac12\left(\frac{g}{m}\right)^2\mu I(\Lambda)
\ee
where $I(\Lambda)$ is defined in (\ref{eq:ILambda}).  The mass renormalization
presented in (\ref{eq:baremass}) then fixes $E^\pm=m$, the physical mass.

We see that, due to the factorization (\ref{eq:factoredpsi}),
the neutral scalars form the coherent state used in Sec.~\ref{sec:single}.  
The location of the source, at $\pm\vec{R}/2$, is encoded
in the phase of the individual wave functions $G_1^\pm$, with the product
in each Fock sector having a phase that corresponds to translation in
light-front coordinates from the origin to $(\pm R_z,\pm\senk{R}/2)$, as
generated by the total momentum.

\section{Double-source expectation value for ${\cal E}$} \label{sec:<E>}

On use of $G_2=0$ and the expression (\ref{eq:G1}) for $G_1^\pm$, the double-source expectation
value $\langle:\!{\cal E}\!:\rangle$ given in (\ref{eq:initial<E>}) becomes
\bea
\langle :\!{\cal E}\!:\rangle&=&\frac{m_0^2}{m}+m \\
&&+\frac12\left(\frac{g}{m}\right)^2\int d\ub{q} \left[\frac{q_\perp^2+\mu^2}{q^+}+q^+\right] 
       \frac{q^+}{16\pi^3}\frac{|e^{iq^+R_z/2+i\senk{q}\cdot\senk{R}/2}+e^{-iq^+R_z/2-i\senk{q}\cdot\senk{R}/2}|^2}
                               {[(q^+)^2+q_\perp^2+\mu^2]^2} \nonumber \\
&&+\frac{g}{2m}\int\frac{d\ub{q}}{\sqrt{16\pi^3 q^+}}\left(-2\frac{g}{m}\right)\sqrt{\frac{q^+}{16\pi^3}}
    \frac{|e^{iq^+R_z/2+i\senk{q}\cdot\senk{R}/2}+e^{-iq^+R_z/2-i\senk{q}\cdot\senk{R}/2}|^2}
                               {(q^+)^2+q_\perp^2+\mu^2}.  \nonumber
\eea
The last two terms differ only in their sign, and a factor of 2, and reduce to
\be
\langle :\!{\cal E}\!:\rangle=\frac{m_0^2}{m}+m
   -\frac12\left(\frac{g}{m}\right)^2\int \frac{d\ub{q}}{16\pi^3} 
       \frac{2+e^{iq^+R_z+i\senk{q}\cdot\senk{R}}+e^{-iq^+R_z-i\senk{q}\cdot\senk{R}}}
                               {(q^+)^2+q_\perp^2+\mu^2}.
\ee
The 2 in the numerator corresponds to a divergent integral that, with a cutoff $\Lambda$,
is proportional to the integral (\ref{eq:ILambda}) previously encountered in the 
single-source case.  From that definition, we have
\be \label{eq:<E>}
\langle :\!{\cal E}\!:\rangle=\frac{m_0^2}{m}+m -\left(\frac{g}{m}\right)^2\mu I(\Lambda)
   -\frac12\left(\frac{g}{m}\right)^2 Y(R),
\ee
with
\be
Y(R)\equiv\int_{q^+>0} \frac{d\ub{q}}{16\pi^3} 
       \frac{e^{iq^+R_z+i\senk{q}\cdot\senk{R}}+e^{-iq^+R_z-i\senk{q}\cdot\senk{R}}}
                               {(q^+)^2+q_\perp^2+\mu^2}
    =\frac12 \int \frac{d\ub{q}}{16\pi^3} 
       \frac{e^{iq^+R_z+i\senk{q}\cdot\senk{R}}+e^{-iq^+R_z-i\senk{q}\cdot\senk{R}}}
                               {(q^+)^2+q_\perp^2+\mu^2}.
\ee
In the second integral there is no restriction on the range of $q^+$.
This unrestricted integral is easily evaluated in spherical coordinates
$\vec{q}=(q_x,q_y,q^+)=(q,\theta,\phi)$, relative to an axis parallel to $\vec{R}$.
The $\phi$ integral is trivial, leaving
\be
Y(R)=\frac{1}{16\pi^2}\int_0^\infty q^2dq\int_{-1}^1 d\cos\theta\frac{e^{iqR\cos\theta}+e^{-iqR\cos\theta}}{q^2+\mu^2}.
\ee
The $\cos\theta$ integral reduces this to
\bea
Y(R)&=&\frac{1}{16\pi^2}\int_0^\infty \frac{q^2 dq}{q^2+\mu^2}
              \left[\frac{e^{iqR}-e^{-iqR}}{iqR}+\frac{e^{-iqR}-e^{iqR}}{-iqR}\right]  \nonumber \\
    &=&\frac{1}{4\pi^2R}\int_0^\infty \frac{q^2 dq}{q^2+\mu^2}\sin(qR)
    =\frac{1}{4\pi^2R}\frac{\pi}{2}e^{-\mu R}.
\eea
Substitution into the expression (\ref{eq:<E>}) for $\langle :\!{\cal E}\!:\rangle$
brings us to a nearly final form
\be
\langle :\!{\cal E}\!:\rangle=\frac{m_0^2}{m}+m -\left(\frac{g}{m}\right)^2\mu I(\Lambda)
   -\left(\frac{g}{2m}\right)^2\frac{e^{-\mu R}}{4\pi R},
\ee
All that remains is to invoke mass renormalization.

\section{Change in number of scalars}  \label{sec:deltan}

The change induced in the number of scalars by the combination of
two static sources is
\be
\langle\delta n\rangle=\int d\ub{q} \left[G_1^{+*}(\ub{q})G_1^-(\ub{q})+G_1^{-*}(\ub{q})G_1^+(\ub{q})\right].
\ee
Substitution of the form (\ref{eq:G1}) for $G_1^\pm$ leaves
\be
\langle\delta n\rangle=\left(\frac{g}{m}\right)^2\int_{q^+>0} d\ub{q} \frac{q^+}{16\pi^3}
             \frac{2+e^{iq^+R_z+i\senk{q}\cdot\senk{R}}+e^{-iq^+R_z-i\senk{q}\cdot\senk{R}}}
                               {[(q^+)^2+q_\perp^2+\mu^2]^2}.
\ee
Use of the same spherical coordinates as in Appendix~\ref{sec:<E>} reduces this to
\be
\langle\delta n\rangle=\left(\frac{g}{m}\right)^2\frac{1}{16\pi^2}
  \int_0^\infty\frac{p^3 dp}{(p^2+\mu^2)^2}\int_{-1}^1\!d\cos\theta\,\cos\theta
           \left[e^{iqR\cos\theta}+e^{-iqR\cos\theta}\right].
\ee
Performance of the $\cos\theta$ integration, and some algebraic rearrangement,
yields
\be
\langle\delta n\rangle=\left(\frac{g}{m}\right)^2\frac{1}{4\pi^2R^2}
     \int_0^\infty \frac{dp}{(p^2+\mu^2)^2}\left[Rp^2\sin(pR)+p\cos(pR)-p\right].
\ee
The individual $p$ integrations can be computed as follows
\be
\int_0^\infty\frac{ p dp}{(p^2+\mu^2)^2}=\frac{1}{2\mu^2},
\ee
\be
\int_0^\infty\frac{p\cos(pR)dp}{(p^2+\mu^2)^2}=\frac{d}{dR}\int_0^\infty\frac{\sin(pR)dp}{(p^2+\mu^2)^2},
\ee
\be
\int_0^\infty\frac{p^2\sin(pR)dp}{(p^2+\mu^2)^2}=-\frac{d}{dR}\int_0^\infty\frac{p\cos(pR)dp}{(p^2+\mu^2)^2},
\ee
\be
\int_0^\infty\frac{\sin(pR)dp}{(p^2+\mu^2)^2}=-\frac{1}{2\mu}\frac{d}{d\mu}\int_0^\infty\frac{\sin(pR)dp}{p^2+\mu^2},
\ee
and~\cite{GR}
\be
\int_0^\infty\frac{\sin(pR)dp}{p^2+\mu^2}=\frac{1}{2\mu}\left[e^{-\mu R}{\rm Ei}(\mu R)-e^{\mu R}{\rm Ei}(-\mu R)\right],
\ee
where ${\rm Ei}$ is the exponential integral function~\cite{DLMF}.  The combination of the
various integrals yields
\be
\langle\delta n\rangle=-\frac{1}{16\pi^2}\left(\frac{g}{m}\right)^2
\left[e^{\mu R}{\rm Ei}(-\mu R)+e^{-\mu R}{\rm Ei}(\mu R)\right].
\ee
%


\end{document}